\documentclass[systeme]{compas2025}

\usepackage{dsfont}
\usepackage{cancel}
\usepackage[title]{appendix}

\toappear{1} 

\begin{document}

\title{Towards Designing an Energy Aware Data Replication Strategy for Cloud Systems Using Reinforcement Learning}
\shorttitle{Eenergy Aware RL-Based Data Replication}

\author{Amir Najjar, Riad Mokadem, Jean-Marc Pierson}

\address{Université de Toulouse\\
Institut de Recherche en Informatique de Toulouse - Campus Rangueil - 
118 Route de Narbonne\\
31062 Toulouse - France\\
\{amir.najjar, riad.mokadem, jean-marc.pierson\}@irit.fr}

\date{\today}

\maketitle


\begin{abstract}
The rapid growth of global data volumes has created a demand for scalable distributed systems that can maintain a high quality of service. Data replication is a widely used technique that provides fault tolerance, improved performance and higher availability. Traditional implementations often rely on threshold-based activation mechanisms, which can vary depending on workload changes and system architecture. System administrators typically bear the responsibility of adjusting these thresholds. To address this challenge, reinforcement learning can be used to dynamically adapt to workload changes and different architectures. In this paper, we propose a novel data replication strategy for cloud systems that employs reinforcement learning to automatically learn system characteristics and adapt to workload changes. The strategy's aim is to provide satisfactory Quality of Service while optimizing a trade-off between provider profit and environmental impact. We present the architecture behind our solution and describe the reinforcement learning model by defining the states, actions and rewards.
\MotsCles{Cloud Systems, Data Replication, Reinforcement Learning, Economic, Energy Consumption.}
\end{abstract}

\section{Introduction}

The growth of social media has generated large data volumes in the cloud. As such, assuring satisfactory Quality of Service (QoS) has become crucial. QoS can be measured using various metrics such as response time (RT), throughput, availability and fault tolerance \cite{armstrong2009towards}. Data replication provides improved RT, fault tolerance and availability through efficient placement of the replicas \cite{goel2007data}.

Previously proposed data replication strategies often utilize thresholds that require prior knowledge of the system and human intervention \cite{lee2012pfrf, bai2013rtrm, tos2016performance}. Statistical methods have been used to predict those thresholds automatically: Khatua et al. \cite{khatua2010optimizing} use time series, Calheiros et al. \cite{calheiros2015workload} use an autoregressive integrated moving average (ARIMA), and Séguéla et al \cite{seguela2022dynamic} use control charts to define such thresholds.

Machine learning (ML) is a promising technique for automatically adapting to system characteristics and adjusting the thresholds without prior knowledge of the underlying architecture or workload. This eliminates the need for human intervention. By using workload traces and environmental state, an ML model can be trained to predict the necessary thresholds for data replication strategies. Additionally, it can use current resource utilization to predict future system performance, aiding in decisions related to activating the data replication mechanism.

Machine learning has been increasingly applied to data replication through supervised learning \cite{bui2016adaptive, shwe2017proactive}, unsupervised learning \cite{sellami2021clustering, symvoulidis2023user}, and reinforcement learning \cite{lu2022rlrp, zhang2024data}. Among these methods, reinforcement learning appears to be the most suitable, as it can learn directly from a simulated or real environment. In contrast, supervised learning relies on labeled datasets, and unsupervised learning requires underlying patterns within the environment \cite{najjar2024review}.

Current strategies, particularly those incorporating machine learning \cite{najjar2024review}, often overlook the economic aspect, which is essential for ensuring profitability in commercial applications such as the cloud \cite{mokadem2020data}. Additionally, they fail to sufficiently address the environmental impact of distributed systems \cite{lindberg2012comparison}. In this paper, we introduce a reinforcement learning-based data replication strategy for cloud systems that balances provider profit with environmental impact in terms of energy consumption and purchase of machines. The initial replication aims to ensure availability and fault tolerance, while the dynamic replication ensures satisfactory response time while optimizing a trade-off between provider profit and energy consumption. VM aggregation is utilized to minimize purchase of additional machines.

The rest of the paper is organized as follows: Section \ref{background} introduces data replication and reinforcement learning. Section \ref{architecture} provides an overview of the considered architecture and its specifications. It defines the necessary terms to introduce our proposed strategy, which is presented in Section \ref{solution}. We describe the reinforcement learning technique by defining the state space, the action space, and the reward signal. Section \ref{conclusion} concludes and presents the next steps required to implement the strategy as well as future plans.

\section{Background}\label{background}

In this section, we provide an overview of data replication and reinforcement learning.

\subsection{Data Replication}

In the context of distributed systems, data replication is the process of creating copies of data in different locations to provide fault tolerance \cite{qu2012rfh}, increased availability \cite{sun2012modeling}, and higher performance \cite{mansouri2016adaptive}. A data replication strategy is required to ensure efficiency of the replication mechanism and to prevent excessive replication. Data replication generally aims to answer the following questions \cite{mokadem2020data}:
\begin{itemize}
    \item Which data to replicate?
    \item What is the activation condition for a replication to occur?
    \item How many replicas (replication factor) are required?
    \item What are the placement locations?
    \item How to minimize the economic cost of replication? This question is particularly relevant for strategies designed for cloud systems.
\end{itemize}

\subsection{Reinforcement Learning}

Machine learning utilizes samples or experience for inference. It can be divided into three methods:

\begin{itemize}
    \item Supervised learning: Characterized by labelled data. The goal is to minimize the loss between the predicted outcome and the true labels. Common tasks are regression and classification.
    \item Unsupervised learning: Characterized by unlabelled data. Unsupervised learning is used to separate the data into groups, to reduce the dimensionality of the data or to learn the relationships in the data.
    \item Reinforcement learning: One or more agents interact with an environment through an action policy. A feedback mechanism known as a reward signal is provided through the agent's exploration of the action space instead of an explicit dataset or labels.
\end{itemize}

Reinforcement learning can be divided into model-based and model-free methods, based on whether the model is known \cite{moerland2023model}. Model-free methods can be used when modeling the environment is challenging. Q-Learning is one such method that utilizes the properties of Markov Decisions Processes (MDP) \cite{puterman2014markov, bellman1952theory} to estimate the reward given a certain state-action pair \cite{sutton2018reinforcement}. With the recent advancements in deep learning \cite{bengio2017deep, lecun2015deep}, its use has been explored in reinforcement learning. Deep Q-Learning \cite{mnih2015human} is a method that employs neural networks to solve the core limitations of Q-Learning: Exploration space explosion and the tendency to overestimate the value of actions.

\begin{figure}
    \centering
    \includegraphics[width=0.9\linewidth]{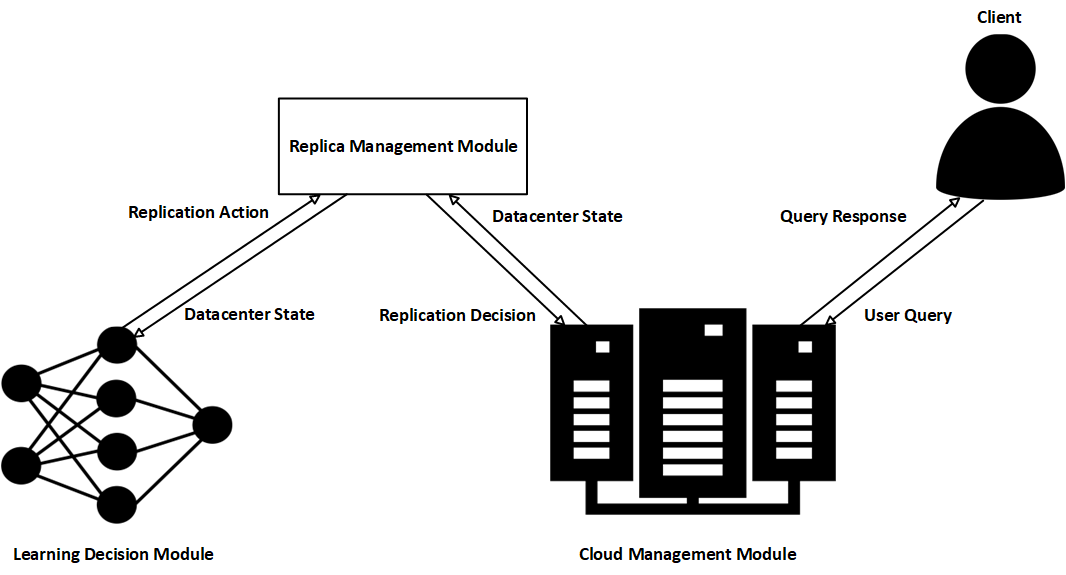}
    \caption{Visual representation of the architecture}
    \label{fig:architecture}
\end{figure}

\section{Architecture}\label{architecture}

The architecture is divided into three main components: The Replica Management Module (RMM), the Learning Decision Module (LDM) and the Cloud Manager Module (CMM). An overarching design of our platform is presented in Figure \ref{fig:architecture}.

\subsection{Overview}

The architecture for our solution is comprised of three main components:

\begin{itemize}
    \item Replica Management Module (RMM): Serves as an intermediary between the LDM and CMM. Replication actions are relayed from the LDM. They are checked for validity and sent as replication decisions to the CMM. Updated system state is relayed from the CMM. It is then batched and sent to the LDM.
    \item Learning Decision Module (LDM): Using updates in the system state, a Deep Q-learning model takes a replication decision and relays it to the RMM.
    \item Cloud Manager Module (CMM): relays static and dynamic information to the RMM. Static information includes datacenter specifications and economic costs (for example, in USD). Dynamic information includes current network and resource utilization, as well as performance metrics such as response time.
\end{itemize}

\subsection{Specification}

The cloud system services a set of $L$ clients $U = \{u_1, ..., u_L\}$ and consists of a set of $N$ datacenters $DC = \{dc_1, ..., dc_N\}$. Each datacenter $dc_i, (1 \leq dc_i \leq N)$ is characterized by a carbon intensity factor of $CI = \begin{bmatrix} ci_i \end{bmatrix}$. Each client $u_l, (1 \leq l \leq L)$ has a certain latency to each of the datacenters $\text{LAT}_l = \{\text{lat}_{l1}, ... \text{lat}_{lN}\}$. Each datacenter has a set of $M_i$ homogeneous hosts $H_i = \{h_{ij}\}, 1 \leq j \leq M_i$. In case a datacenter has heterogeneous hosts, it can be considered as multiple homogeneous datacenters sharing a common backbone. Hosts of $dc_i$ are characterized by a CPU with $C = \begin{bmatrix} c_i \end{bmatrix}$ cores, a core frequency of $F = \begin{bmatrix} f_i \end{bmatrix}$, and economic cost per execution cycle of $CC = \begin{bmatrix} cc_i \end{bmatrix}$. Each host has $S = \begin{bmatrix} s_i \end{bmatrix}$ bytes of storage capacity and a storage cost of $SC = \begin{bmatrix} sc_i \end{bmatrix}$ per byte. Several VMs can be deployed on a host depending on the core count and the required resources. A VM on $h_{ij}$ is denoted by $v_{ijk}$. A VM $v_{ijk}$ can occupy one or more cores of the host $h_{ij}$. The number of cores occupied by $v_{ijk}$ is referred to by $nco_{ijk}$.

\begin{figure}
    \centering
    \includegraphics[width=1.0\linewidth]{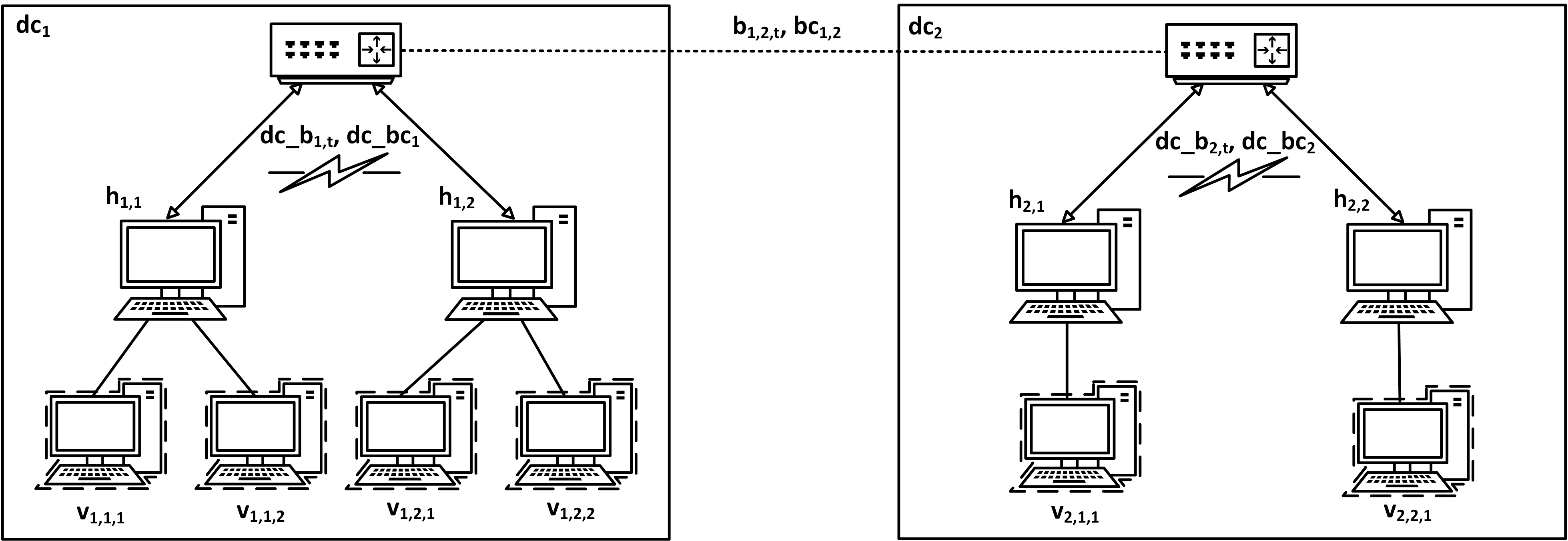}
    \caption{Architecture example}
    \label{fig:example}
\end{figure}

Network links are established between all pairs of datacenters: for $dc_i$ and $dc_{i'} \ i \neq i'$, $b_{ii', t}$ denotes the bandwidth available at time $t$, and $bc_{ii'}$ denotes the bandwidth economic cost per byte. Similarly, $dc\_b_{i,t}$ denotes the bandwidth available at time $t$ for the backbone network of $dc_i$, and $dc\_bc_{i}$ denotes the bandwidth economic cost per byte. It is possible to represent the network properties in four matrices, $B_t = \begin{bmatrix} b_{ii', t} \end{bmatrix}$, $BC = \begin{bmatrix} bc_{ii'} \end{bmatrix}$, $DC\_B_t = [dc\_b_{i, t}]$, $DC\_BC = [dc\_bc_i]$, which will be relevant for the solution design. Figure \ref{fig:example} presents an example of a platform with two datacenters. $dc_1$ has two hosts, each of which has two VMs. $dc_2$ has two hosts with one VM each.

A Service Level Agreement (SLA) is established between the provider and a client $u_l$. The client pays a fixed rate per query $RATE_l$. Multiple Service Level Objectives (SLO) are defined: 
\begin{itemize}
    \item Availability objective $AVO_l$: Minimum replication factor to respect at all times,
    \item Response time objective $RTO_l$: Maximum allowed response time before incurring a penalty $RT\_PEN_l$.
\end{itemize}

For each datum $d_l$ of size $sz_l$ associated to user $u_l$, we retain visibility of the replication factor in each datacenter $R_l = [ r_{l1}, ..., r_{lN} ]$ as well as the total number of replicas across the platform $GR_l = \sum_{i=1}^{N} r_{li} \geq AVO_l$.

A client $u_l$ communicates with the RMM to request $d_l$. The query arrival rate follows a Poisson process \cite{cao2003internet, xiong2015smart}. A query $Q$ at time $t$ is characterized by an arrival time $t(Q)$ and an amount of execution cycles $ec(Q)$. At time $t$, we also calculate the average CPU load percentage of hosts containing $d_l$ per datacenter $UTIL_{l, t} = \begin{bmatrix} util_{l1,t}, ..., util_{lN, t} \end{bmatrix}$. The definition of each term can be found in appendix A.

\section{Proposed Strategy}\label{solution}

To avoid the environmental impact incurred by purchasing more machines, we adopt VM aggregation; data is replicated on hosts that are already active to minimize the amount of active hosts. We describe the strategy's approach to initial placement as well as dynamic allocation.

\subsection{Initial Replication}
To satisfy the availability objective for each user, $AVO_l$ replicas need to be created. The RMM places the initial copies in different datacenters to ensure fault tolerance. The initial datum is placed in the datacenter with the lowest latency to the user to ensure fast response time. The rest of the copies are placed in the datacenters with the lowest carbon intensity $ci_i$. Response time is not as crucial for the remaining copies as they are used as backup in case a fault related to the VM containing the initial datum occurs.

\subsection{Dynamic Replication}

The LDM uses a Deep Q-Learning neural network to take actions based on a state. The state is encoded in the input layer and the value of actions is encoded in the output layer. The state consists of:

\begin{itemize}
    \item Datacenter information: $F$ core frequencies, $C$ core counts, $CC$ costs per execution cycle, $SC$ storage costs, carbon intensities $CI$,
    \item Network information: $B_t$ inter-DC bandwidth available at time $t$, $BC$ inter-DC bandwidth costs, $DC\_B_t$ intra-DC bandwidth available at time $t$, $DC\_BC$ intra-DC bandwidth costs,
    \item User information: $\text{LAT}_{l}$ latencies, $UTIL_{l, t}$ average utilization, $sz_l$ data size, $RTO_l$ RT objective.
\end{itemize}

A preliminary check is performed before providing the state space. For example, in case of insufficient cores or disk space in a certain data center, the frequency of said data center is set to 0 in the state space. Furthermore, replication action is verified for validity and transformed into a replication decision. State-action-reward tuples are batched periodically. However, upon incurring penalties, batching is invoked and a replication decision is taken to remedy the response time loss.

The action space consists of:
\begin{itemize}
    \item replicating in a datacenter $dc_i$,
    \item not performing any replication
\end{itemize}
In case of an invalid action, the action with the next best value is taken into consideration until a valid action is reached. 

The reward signal is constructed from economic profit and a penalty on energy consumption.

\begin{equation}\label{eq:reward}
    Reward(Q) = \alpha \cdot Economic(Q) - \beta \cdot Energy(Q)
\end{equation}

Where $\alpha, \beta \geq 0$ are parameters set by the provider based on how much they're willing to compromise profit for saving energy. The economic part is the difference between the rate per user query and possible penalty cost in addition to the cost of utilized resources which are comprised of the CPU cost, the storage cost, bandwidth cost, and the replication cost. It is defined as follows:

\begin{equation}\label{eq:econ}
    Economic(Q) = RATE_l - (CPU(Q) + STOR(Q) + BW(Q) + PEN(Q))
\end{equation}

The CPU cost can be calculated using the execution cycles $ec(Q)$, the core count $nco_{ijk}$ and the cost per execution cycle $cc_i$. For simplicity, we assume that task performance scales linearly with core count:

\begin{equation}
    CPU(Q) = \dfrac{ec(Q)}{\cancel{nco_{ijk}}} \cdot cc_i \cdot \cancel{nco_{ijk}} = ec(Q) \cdot cc_i
\end{equation}

The storage is the size of the datum $sz_l$ multiplied by the sum of storage costs incurred in each datacenter, which can be calculated using the product of the replication factor $r_{li}$ and the storage cost $sc_i$:

\begin{equation}
    STOR(Q) = sz_l \cdot \sum_{i=1}^{N} sc_{i} \cdot r_{li}
\end{equation}

Replication can occur between two hosts in the same datacenter $(h_{ij} \rightarrow h_{ij'}, j \neq j')$ or between two hosts in different datacenters $(h_{ij} \rightarrow v_{i'j'}, i \neq i')$, the bandwidth cost associated is calculated accordingly:
\begin{equation}
    BW(Q) = 
    \begin{cases}
        \hfil sz_l \cdot dc\_bc_i & \text{in same DC}\\
        \hfil sz_l \cdot bc_{ii'} & \text{in different DCs }\\
        \hfil 0 & \text{no replication}
    \end{cases}
\end{equation}

The final part of equation \ref{eq:econ} is the penalty $RT\_PEN_l$ incurred if $RTO_l$ is not respected:

\begin{equation}
    PEN(Q) =
    \begin{cases}
        \hfil RT\_PEN_l & RT(Q) > RTO_l\\
        \hfil 0 & RT(Q) \leq RTO_l
    \end{cases}
\end{equation}

The energy part of the reward signal in equation \ref{eq:reward} is defined as follows:

\begin{equation}
    Energy(Q) = p_{ijk, F(Q)} - p_{ijk,t(Q)} + p_{REPL, Q}
\end{equation}

Where:

\begin{itemize}
    \item $t(Q), F(Q)$ are the arrival time and the finish time of query $Q$ respectively
    \item $p_{ijk,t}$ is the power consumption of $v_{ijk}$ executing $Q$ at time $t$
    \item $p_{REPL, Q}$ is the energy consumed to create a replica if a replication occurred
\end{itemize}

\section{Conclusion and Future Work}\label{conclusion}
We have explored the possibility of designing a data replication strategy based on reinforcement learning in order to respond to the general lack in the literature for strategies that take into account the economic and environmental aspect. Reinforcement learning also allows us to eliminate human intervention commonly required in strategy design. We have presented the architecture behind our solution and defined the solution space as well as the reward signal.

Our next steps are: designing a suitable neural network layout used, implementing further optimizations for Deep Q-Learning such as Experience Replay \cite{schaul2015prioritized} and normalizing the components of the reward signal to facilitate the choice of the weighted parameters for the provider. In addition, we plan to consider the energy consumption of the learning decision module and the cost of its usage. Further in the future, we plan to test our strategy in both simulated environments and testbeds. In addition, we plan to compare against other state of the art strategies, notably ones that utilize machine learning in their design. 

\bibliography{references}

\newpage

\appendix

\textbf{Appendix A}

\begin{table}[h]
    \centering
    \begin{tabular}{|l|l|}
        \hline
         Term & Definition \\
         \hline
         L & Number of clients \\
         \hline
         $U, u_l$ & Client \\
         \hline
         $d_l$ & Datum \\
         \hline
         $N$ & Number of datacenters \\
         \hline
         $DC$, $dc_i$ & Datacenter \\
         \hline
         $CI, ci_i$ & Carbon intensity \\
         \hline
         $\text{LAT}_l, \text{lat}_{li}$ & User latency \\
         \hline
         $M_i$ & Number of hosts of $dc_i$ \\
         \hline
         $H_i, h_{ij}$ & Host \\
         \hline
         $C, c_i$ & CPU core count \\
         \hline
         $CC, cc_i$ & CPU cost per unit of time \\
         \hline
         $S, s_i$ & Storage capacity \\
         \hline
         $SC, sc_i$ & Storage cost per byte \\
         \hline
         $v_{ijk}$ & VM \\
         \hline
         $nco_{ijk}$ & Number of cores occupied by VM \\
         \hline
         $B_t, b_{ii', t}$ & Inter-DC bandwidth \\
         \hline
         $BC, bc_{ii'}$ & Inter-DC bandwidth cost per byte \\
         \hline
         $DC\_B_t, dc\_b_{i, t}$ & Intra-DC bandwidth \\
         \hline
         $DC\_BC, dc\_bc_{ij}$ & Intra-DC bandwidth cost per byte \\ 
         \hline
         $RATE_l$ & Rate per query \\
         \hline
         $AVO_l$ & Availability objective \\ 
         \hline
         $RTO_l$ & Response time objective \\
         \hline
         $RT\_PEN_l$ & Penalty incurred upon failing to meet $RTO_l$ \\
         \hline
         $sz_l$ & Size of $d_l$ in bytes \\
         \hline
         $R_l, r_{li}$ & Replication factors \\
         \hline
         $GR_l$ & Global replication factor of $d_l$ \\
         \hline
         $Q$ & Query \\
         \hline
         $t(Q)$ & Time of arrival \\
         \hline
         $ec(Q)$ & Amount of execution cycles \\
         \hline
         $UTIL_{l, t}, util_{li, t}$ & Average utility in $dc_i$ \\
         \hline         
    \end{tabular}
    \caption{Terms and definitions}
    \label{tab:terms}
\end{table}

\end{document}